\documentclass[onecolumn,superscriptaddress,secnumarabic,nobibnotes,aps,prd,nofootinbib,altaffilletter,11pt]{revtex4}
\usepackage{amsmath,amssymb}
\usepackage{graphicx}
\usepackage{dcolumn}
\usepackage{bm,color}
\usepackage{hyperref}
\usepackage{accents}
\usepackage{amssymb,float}
\usepackage{amsmath}
\usepackage{multirow}
\usepackage{siunitx}
\usepackage{tabularx}
\usepackage{booktabs}
\usepackage{url}
\usepackage{bm}
\usepackage{braket}
\usepackage{tikz}
\usepackage{url}
\usepackage{accents}
\usepackage{natbib}
\usepackage{subfigure}

\usetikzlibrary{positioning}

\graphicspath{ {figs/} }
\hypersetup{colorlinks=true,citecolor=blue, linkcolor=red, urlcolor=blue}

\begin{document}

\title{Noether Symmetries in $f\left( Q\right) -$Cosmology}
\author{Konstantinos F. Dialektopoulos}
\email{kdialekt@gmail.com}
\affiliation{Department of Mathematics and Computer Science, Transilvania University of
Brasov, Eroilor 29, Brasov, Romania}
\author{Genly Leon }
\email{genly.leon@ucn.cl}
\affiliation{Departamento de Matem\'{a}ticas, Universidad Cat\'{o}lica del Norte, Avenida
Angamos 0610, Casilla 1280 Antofagasta, Chile}
\affiliation{Institute of Systems Science, Durban University of Technology, Durban 4000, South Africa}
\author{Andronikos Paliathanasis}
\email{anpaliat@phys.uoa.gr}
\affiliation{Institute of Systems Science, Durban University of Technology, Durban 4000, South Africa}
\affiliation{School for Data Science and Computational Thinking, Stellenbosch University, 44 Banghoek Rd, Stellenbosch 7600, South Africa}
\affiliation{Departamento de Matem\'{a}ticas, Universidad Cat\'{o}lica del Norte, Avenida
Angamos 0610, Casilla 1280 Antofagasta, Chile}

\begin{abstract}
We apply the Noether symmetry analysis in $f\left( Q\right)$-Cosmology to
determine invariant functions and conservation laws for the cosmological
field equations. For the FLRW background and the four families of
connections, it is found that only power-law $f\left( Q\right)$ functions
admit point Noether symmetries. Finally, exact and analytic solutions are
derived using the invariant functions.
\end{abstract}

\maketitle

\section{Introduction}

\label{sec:intro}

Symmetric teleparallel theory of gravity \cite{Nester:1998mp} and its
generalizations \cite{fq1,fq2,fq3,fq5,sc1,gg1,gg2,gg22,gg3,gg4,gg5} have
recently drawn the attention of cosmologists, as it provides a new geometric
approach for describing cosmic acceleration and cosmic structure. In symmetric teleparallel theory, the geometry of the physical metric is
non-Riemannian, and the autoparallels are independent of the metric tensor.
Specifically, the connection is defined to be symmetric and flat, differing
from the Levi-Civita connection. Thus, the connection introduces
geometrodynamical degrees of freedom in the gravitational model. These new
degrees of freedom lead to novel behaviors in the evolution of physical
parameters, allowing for new physics. Cosmological applications of symmetric
teleparallel theory, where dark energy is interpreted as a geometric
phenomenon, can be found in \cite{cc1,cc2,cc3,cc4,cc5,cc6,cc7,cc8,cc9}.
Additionally, studies on compact astrophysical objects are presented in \cite%
{cc10,cc11,cc12,cc13,cc14}. For a recent review of symmetric teleparallel
theory, we refer the reader to \cite{revh}.

In this work, we focus on studying the cosmological field equations in $%
f\left( Q\right)$-gravity \cite{fq1,fq2}. In symmetric teleparallel $f\left(
Q\right)$-gravity, the gravitational Action Integral is defined by an
arbitrary function $f$ of the nonmetricity scalar $Q$. When $f$ is a linear
function, the Symmetric Teleparallel Equivalent to General Relativity
(STEGR) theory is recovered \cite{Nester:1998mp}. This gravitational theory
belongs to the same family of nonlinear $f$-theories \cite%
{fr1,fr2,fr3,ft1,ft2} proposed for the Ricci scalar and the torsion scalar. For an extensive review on $f(R)$ theories, check \cite{sot1,sot2}, while for $f(T)$ theory, check \cite{Bahamonde:2021gfp}.
The geometrodynamical degrees of freedom in $f\left(Q\right)$-gravity can be
associated with scalar fields \cite{mini}, and the field equations admit a
minisuperspace description. Consequently, the method of variational
symmetries can be applied to analyze the integrability properties of the
field equations.

The Noether symmetry analysis is a powerful technique for determining
conservation laws for nonlinear differential equations derived from a
variational principle. Specifically, we identify the function $%
f\left(Q\right)$ for which the field equations admit variational symmetries.
Using Noether's theorem, we construct the conservation laws admitted by the
dynamical system. This classification approach has been extensively studied
in modified theories of gravity \cite%
{ns1,ns2,ns3,ns5,ns6,ns7,ns8,ns9,ns10,ns11}. As we shall discuss, the
Noether symmetry analysis serves as a geometric selection rule for choosing the free parameters and functions within gravitational models \cite{anrev1,Dialektopoulos:2018qoe}.

A partial analysis of the Noether symmetries in $f\left( Q\right) $-gravity within the Friedmann--Lema\^{\i}tre--Robertson--Walker (FLRW)\ geometry performed in \cite{fqns}. Nevertheless, in this study we performed a detailed analysis of the classification problem. We show that the Noether symmetry analysis can used for the derivation of scaling solutions and for the derivation of an integrable cosmological model within a spatially flat FLRW geometry. 

It is known in the literature that $f(Q)$ theory is plagued with ghosts around Minkowski and FLRW spacetimes \cite{Gomes:2023tur}. Despite that, it remains an intriguing alternative to General Relativity because it provides a geometrically novel approach to gravity based purely on non-metricity, avoiding curvature and torsion. Additionally, specific subclasses or extensions of the theory may offer viable cosmological models that explain dark energy or late-time acceleration without requiring a cosmological constant. Moreover, studying $f(Q)$ gravity helps to explore the broader landscape of metric-affine theories and deepen our understanding of how different geometric structures influence gravitational dynamics. Even if the full theory is problematic, certain well-behaved limits or modified versions could lead to new insights into quantum gravity or alternative formulations of gravity beyond General Relativity. For more details we refer to the discussion in \cite{guzman}.

The structure of the paper is as follows: In Section \ref{sec2}, we briefly discuss the basic definitions of symmetric
teleparallel gravity and its generalization, $f\left(Q\right)$-gravity. For
the background space, we consider an FLRW metric and present the
corresponding field equations. The mathematical framework of the Noether
symmetry analysis is introduced in Section \ref{sec3}, where we discuss the
application of Noether's theorem as a geometric selection rule in $%
f\left(Q\right)$-gravity. In Section \ref{sec4}, we present the symmetry classification scheme for the
cosmological Lagrangian of $f\left(Q\right)$-gravity and utilize the Noether
symmetries to derive conservation laws. In Section \ref{sec5}, we apply
these results to determine exact and analytic solutions. Finally, in Section %
\ref{sec6}, we summarize our findings and draw conclusions.

\section{Symmetric teleparallel geometry}

\label{sec2}

The geometric construction of the symmetric teleparallel theory is that the
physical space is described by a symmetric second-rank metric tensor $g_{\mu
\nu }$, while the autoparallels are defined according to the symmetric and
flat connection $\Gamma _{\mu \nu }^{\lambda }$, with convariant derivative $%
\nabla $ which is different from the Levi-Civita connection. By definition,
it the Riemann tensor $R_{\;\lambda \mu \nu }^{\kappa }\left( \Gamma _{\mu
\nu }^{\lambda }\right) $ and the torsion tensor $\mathrm{T}_{\;\mu \nu
}^{\lambda }\left( \Gamma _{\mu \nu }^{\lambda }\right) $ are zero, that is 
\cite{revh}%
\begin{gather}
R_{\;\lambda \mu \nu }^{\kappa }\left( \Gamma _{\mu \nu }^{\lambda }\right)  = 2\partial_{[\mu}\Gamma^{\alpha}_{\;\;\nu]\beta}+2\Gamma^{\alpha}_{\;\;[\mu|\lambda|}\Gamma^{\lambda}_{\;\;\nu]\beta} = 0\,,  \label{fq.01} \\
\mathrm{T}_{\;\mu \nu }^{\lambda }\left( \Gamma _{\mu \nu }^{\lambda }\right)  = \Gamma ^\lambda {}_{\mu\nu} - \Gamma^\lambda {}_{\nu\mu} = 0\,.  \label{fq.02}
\end{gather}

The gravitational Action Integral of STEGR is \cite{Nester:1998mp}%
\begin{equation}
S_{Q}=\int d^{4}x\sqrt{-g}Q+S_{m},  \label{fq.03}
\end{equation}%
where $S_{m}=\int d^{4}x\sqrt{-g}L_{m}$ describes the Action Integral for
the matter components with matter Lagrangian $L_{m}$. $Q$\ is the nonmetricity scalar which is the fundamental scalar in
symmetric teleparallel theory of gravity defined as \cite{revh} 
\begin{equation}
Q\left( \Gamma _{\mu \nu }^{\lambda }\right) =Q_{\lambda \mu \nu }P^{\lambda
\mu \nu },  \label{fq.04}
\end{equation}%
where $Q_{\lambda \mu \nu }$ is the nonmetricity tensor, 
\begin{equation}
Q_{\lambda \mu \nu }= \nabla _{\lambda
}g_{\mu \nu } = g_{\mu \nu ,\lambda }-\Gamma _{\;\lambda \mu }^{\sigma
}g_{\sigma \nu }-\Gamma _{\;\lambda \nu }^{\sigma }g_{\mu \sigma },
\label{fq.05}
\end{equation}%
and%
\begin{equation}
P_{~\mu \nu }^{\lambda }=\frac{1}{4}\left( -2L_{~~\mu \nu }^{\lambda
}+Q^{\lambda }g_{\mu \nu }-Q^{\prime \lambda }g_{\mu \nu }-\delta _{(\mu
}^{\lambda }Q_{\nu )}\right) ,  \label{fq.06}
\end{equation}%
is the nonmetricity conjugate tensor, while $L_{~\mu \nu }^{\lambda }$,~$%
Q_{\lambda }$ and $Q_{\lambda }^{\prime }$ are defined as \cite{revh} 
\begin{equation}
L_{~\mu \nu }^{\lambda }=\frac{1}{2}g^{\lambda \sigma }\left( Q_{\mu \nu
\sigma }+Q_{\nu \mu \sigma }-Q_{\sigma \mu \nu }\right) ,  \label{fq.07}
\end{equation}%
and%
\begin{equation}
Q_{\lambda }=Q_{\lambda ~~~\mu }^{~~~\mu },Q_{\lambda }^{\prime
}=Q_{~~\lambda \mu }^{\mu }~.  \label{fq.08}
\end{equation}

By definition the nonmetricity scalar $Q$ for the connection $\Gamma _{\mu
\nu }^{\lambda }$, and the Ricci scalar for the Levi-Civita connection of
the metric tensor $g_{\mu \nu }$, are related as \cite{revh}%
\begin{equation}
Q=R+B,  \label{fq.09}
\end{equation}%
in which $B$ is a boundary term. \ 

Consequently, the variation of the Action Integral (\ref{fq.03}) leads to
the same filed equations with that of the Einstein-Hilbert Action Integral
of General Relativity.

\subsection{$f\left( Q\right) -$gravity}

The introduction of nonlinear components of the nonmetricity scalar within the Action Integral leads to a gravitational theory different from General Relativity. This family of models form the $f\left( Q\right) $-gravity, with
Action Integral \cite{fq1,fq2}%
\begin{equation}
S_{f\left( Q\right) }=\int d^{4}x\sqrt{-g}f\left( Q\right) +S_{m},
\label{fq.10}
\end{equation}%
where in the following we assume that $f$ is a nonlinear function. Because
of the condition (\ref{fq.09}), $f\left(Q\right) $-gravity is different from the $f\left(R\right) $-gravity.

Variation of (\ref{fq.10}) with respect to the metric tensor leads to the field equations 
\begin{equation}
\frac{2}{\sqrt{-g}}\nabla _{\lambda }\left( \sqrt{-g}f^{\prime }(Q)P_{\;\mu
\nu }^{\lambda }\right) -\frac{1}{2}f(Q)g_{\mu \nu }+f^{\prime }(Q)\left( P_{\mu \rho \sigma }Q_{\nu }^{\;\rho \sigma }-2Q_{\rho \sigma \mu }P_{%
\phantom{\rho\sigma}\nu }^{\rho \sigma }\right) =T_{\mu \nu },  \label{fq.11}
\end{equation}%
or in a similar-to-GR form,
\begin{equation}
f^{\prime }(Q)G_{\mu \nu }+\frac{1}{2}g_{\mu \nu }\left( f^{\prime
}(Q)Q-f(Q)\right) +2f^{\prime \prime }(Q)\left( \nabla _{\lambda }Q\right)
P_{\;\mu \nu }^{\lambda }=T_{\mu \nu },  \label{fq.12}
\end{equation}%
where now a prime denotes total derivative with respect to the nonmetricity
scalar $Q$, i.e. $f^{\prime }(Q)=\frac{df\left( Q\right) }{dQ}$ and $G_{\mu
\nu }$ is the Einstein tensor. Tensor field $T_{\mu \nu }$ describes the
contribution of the matter source in the gravitational field equations. For
the matter source, we assume that it is self-consistent, that is, $%
T_{~~~;\nu }^{\mu \nu }=0$, in which $";"$ remark the covariant derivative with respect to the Levi-Civita connection. Due to the latter condition, variation of the Action Integral (\ref{fq.10})
with respect to the connection leads to the equation of motion for the
connection 
\begin{equation}
\nabla _{\mu }\nabla _{\nu }\left( \sqrt{-g}f^{\prime }(Q)P_{\phantom{\mu\nu}%
\sigma }^{\mu \nu }\right) =0.  \label{fq.13}
\end{equation}%
This equation describes the evolution of the geometrodynamical degrees of
freedom introduced in the field equations by the connection.

Since the connection is flat, there exists a coordinate system for which $\Gamma _{\mu \nu }^{\lambda }=0$ and equation (\ref{fq.13}) is trivially satisfied. This system is known as the coincidence gauge. However, for an arbitrary coordinate system, the connection equation (\ref{fq.13}) is
not trivially satisfied. In the latter system, the number of the dynamical degrees of freedom is increased. We should clarify that we refer to the degrees of freedom of the differential equations. 

This will be clarified in the following lines where we will discuss the selection of the
connection in the case of a FLRW cosmology.

\subsection{FLRW Cosmology}

Let us assume that the universe is described by the FLRW geometry with line
element 
\begin{equation}
ds^{2}=-N(t)^{2}dt^{2}+a(t)^{2}\left[ \frac{dr^{2}}{1-kr^{2}}+r^{2}\left(
d\theta ^{2}+\sin ^{2}\theta d\varphi ^{2}\right) \right] ,  \label{fq.14}
\end{equation}%
where $N\left(t\right) $ is the lapse function, $a\left( t\right) $ is the
scale factor and $k$ is the spatial curvature. For $k=0$, the spacetime is spatially flat, for $k=1$ it is closed, and for $k=-1$ it is open. Consider $%
u^{\mu }=\frac{1}{N}\delta _{t}^{\mu }$, to be the comoving observer, then
the expansion rate is defined as $\Theta =3H$, where $H=\frac{1}{N}\frac{%
\dot{a}}{a}$ is the Hubble function and $\dot{a}=\frac{da}{dt}$.

The matter source is assumed to be dust, related to the dark matter of universe and its Lagrangian is defined as 
\begin{equation}
L_{m}=\rho _{m0}a^{-3}.
\end{equation}

The definition of a symmetric and flat connection for the FLRW spacetime (\ref{fq.14}) is not unique. Even though, the selection of the connection is not essential in STEGR, in $f\left( Q\right) $-gravity the choice of the connection plays an important role.

There are four different connections which are symmetric and flat and lead
to four different definitions for the nonmetricity scalar $Q$. Specifically,
three connections are for the spatially flat universe, i.e. $k=0$, and the
fourth connection is for $k=\pm 1$~\cite{he1,he2,he3}.

For the spatially flat FLRW spacetime in the coordinates described by the
line element (\ref{fq.14}) the three different connections have the common
components, 
\begin{equation*}
\Gamma _{\theta \theta }^{r}=-r~,~\Gamma _{\varphi \varphi }^{r}=-r\sin
^{2}\theta
\end{equation*}%
\begin{equation*}
\Gamma _{\varphi \varphi }^{\theta }=-\sin \theta \cos \theta ~,~\Gamma
_{\theta \varphi }^{\varphi }=\Gamma _{\varphi \theta }^{\varphi }=\cot
\theta
\end{equation*}%
\begin{equation*}
\Gamma _{\;r\theta }^{\theta }=\Gamma _{\;\theta r}^{\theta }=\Gamma
_{\;r\varphi }^{\varphi }=\Gamma _{\;\varphi r}^{\varphi }=\frac{1}{r},
\end{equation*}%
which are nothing else, than the Levi-Civita connection for the
three-dimensional flat space 
\begin{equation}
ds_{\left( 3\right) }^{2}=dr^{2}+r^{2}\left( d\theta ^{2}+\sin ^{2}\theta
d\varphi ^{2}\right) ,
\end{equation}%
expressed in spherical coordinates.

Connection $\Gamma ^{A}$ has the additional nonzero components%
\begin{equation}
\Gamma _{\;tt}^{t}=\gamma (t),  \label{con1}
\end{equation}%
where $\gamma (t)$ is a function of the time variable $t$. The nonmetricity
scalar is defined \cite{he1,he2,he3}%
\begin{equation*}
Q\left( \Gamma ^{A}\right) =-6H^{2}\text{.}
\end{equation*}%
Hence, the resulting field
equations are%
\begin{gather}
3H^{2}f^{\prime }(Q)+\frac{1}{2}\left( f(Q)-Qf^{\prime }(Q)\right) =\rho
_{m0}a^{-3} \\
-\frac{2}{N}\frac{d}{dt}\left( f^{\prime }\left( Q\right) H\right)
-3H^{2}f^{\prime }(Q)-\frac{1}{2}\left( f(Q)-Qf^{\prime }(Q)\right) =0.
\end{gather}%
Indeed, the definition of the scalar $\gamma \left( t\right) $ plays no role
in the gravitational dynamics, that is, equation (\ref{fq.13}) is trivially
satisfied.

Connection $\Gamma ^{B}$ has the additional nonzero components 
\begin{equation*}
\Gamma _{\;tt}^{t}=\frac{\dot{\gamma}(t)}{\gamma (t)}+\gamma (t),\quad
\Gamma _{\;tr}^{r}=\Gamma _{\;rt}^{r}=\Gamma _{\;t\theta }^{\theta }=\Gamma
_{\;\theta t}^{\theta }=\Gamma _{\;t\varphi }^{\varphi }=\Gamma _{\;\varphi
t}^{\varphi }=\gamma (t).
\end{equation*}%
where now the nonmetricity scalar is defined as \cite{he1,he2,he3}%
\begin{equation}
Q\left( \Gamma ^{B}\right) =-6H^{2}+\frac{3\gamma }{N}\left( 3H-\frac{\dot{N}%
}{N^{2}}\right) +\frac{3\dot{\gamma}}{N^{2}},
\end{equation}%
and the field equations are%
\begin{gather}
3H^{2}f^{\prime }(Q)+\frac{1}{2}\left( f(Q)-Qf^{\prime }(Q)\right) +\frac{%
3\gamma \dot{Q}f^{\prime \prime }(Q)}{2N^{2}}  = \rho _{m0}a^{-3}, \\
-\frac{2}{N}\frac{d}{dt}\left( f^{\prime }(Q)H\right) -3H^{2}f^{\prime }(Q)-%
\frac{1}{2}\left( f(Q)-Qf^{\prime }(Q)\right) +\frac{3\gamma \dot{Q}%
f^{\prime \prime }(Q)}{2N^{2}}  = 0, \\
\dot{Q}^{2}f^{\prime \prime \prime }(Q)+\left[ \ddot{Q}+\dot{Q}\left( 3NH-%
\frac{\dot{N}}{N}\right) \right] f^{\prime \prime }(Q) = 0.
\end{gather}%

Connection $\Gamma ^{C}$ is described by the nonzero components%
\begin{equation*}
\Gamma _{\;tt}^{t}=-\frac{\dot{\gamma}(t)}{\gamma (t)},\quad \Gamma
_{\;rr}^{t}=\gamma (t),\quad \Gamma _{\;\theta \theta }^{t}=\gamma
(t)r^{2},\quad \Gamma _{\;\varphi \varphi }^{t}=\gamma (t)r^{2}\sin
^{2}\theta\,,
\end{equation*}%
for which, the nonmetricity scalar reads 
\begin{equation*}
Q\left( \Gamma ^{C}\right) =-\frac{6\dot{a}^{2}}{N^{2}a^{2}}+\frac{3\gamma }{%
a^{2}}\left( \frac{\dot{a}}{a}+\frac{\dot{N}}{N}\right) +\frac{3\dot{\gamma}%
}{a^{2}},
\end{equation*}%
while the gravitational field equations are%
\begin{gather}
3H^{2}f^{\prime }(Q)+\frac{1}{2}\left( f(Q)-Qf^{\prime }(Q)\right) -\frac{%
3\gamma \dot{Q}f^{\prime \prime }(Q)}{2a^{2}}  = \rho _{m0}a^{-3}, \\
-\frac{2}{N}\frac{d}{dt}\left( \frac{f^{\prime }(Q)\dot{a}}{Na}\right) -%
\frac{3\dot{a}^{2}}{N^{2}a^{2}}f^{\prime }(Q)-\frac{1}{2}\left(
f(Q)-Qf^{\prime }(Q)\right) +\frac{\gamma \dot{Q}f^{\prime \prime }(Q)}{%
2a^{2}} = 0, \\
\dot{Q}^{2}f^{\prime \prime \prime }(Q)+\left[ \ddot{Q}+\dot{Q}\left( \frac{%
\dot{a}}{a}+\frac{\dot{N}}{N}+\frac{2\dot{\gamma}}{\gamma }\right) \right]
f^{\prime \prime }(Q)  = 0.
\end{gather}

Finally, for the case where $k=\pm 1$, connection $\Gamma ^{D}$ has the
nonzero components%
\begin{equation}
\begin{split}
& \Gamma _{\;tr}^{r}=\Gamma _{\;rt}^{r}=\Gamma _{\;t\theta }^{\theta
}=\Gamma _{\;\theta t}^{\theta }=\Gamma _{\;t\varphi }^{\varphi }=\Gamma
_{\;\varphi t}^{\varphi }=-\frac{k}{\gamma (t)},\quad \Gamma _{\;rr}^{r}=%
\frac{kr}{1-kr^{2}}, \\
& \Gamma _{\;\theta \theta }^{r}=-r\left( 1-kr^{2}\right) ,\quad \Gamma
_{\;\varphi \varphi }^{r}=-r\sin ^{2}(\theta )\left( 1-kr^{2}\right) \quad
\Gamma _{\;r\theta }^{\theta }=\Gamma _{\;\theta r}^{\theta }=\Gamma
_{\;r\varphi }^{\varphi }=\Gamma _{\;\varphi r}^{\varphi }=\frac{1}{r}, \\
& \Gamma _{\;\varphi \varphi }^{\theta }=-\sin \theta \cos \theta ,\quad
\Gamma _{\;\theta \varphi }^{\varphi }=\Gamma _{\;\varphi \theta }^{\varphi
}=\cot \theta ,
\end{split}%
\end{equation}%
and%
\begin{equation*}
\Gamma _{\;tt}^{t}=-\frac{k+\dot{\gamma}(t)}{\gamma (t)},\quad \Gamma
_{\;rr}^{t}=\frac{\gamma (t)}{1-kr^{2}}\quad \Gamma _{\;\theta \theta
}^{t}=\gamma (t)r^{2},\quad \Gamma _{\;\varphi \varphi }^{t}=\gamma
(t)r^{2}\sin ^{2}(\theta ).
\end{equation*}%
The nonmetricity scalar reads 
\begin{equation}
Q\left( \Gamma ^{D}\right) =-6H^{2}+\frac{3\gamma }{a^{2}}N\left( H+\frac{%
\dot{N}}{N^{2}}\right) +\frac{3\dot{\gamma}}{a^{2}}+k\left[ \frac{6}{a^{2}}+%
\frac{3}{\gamma N^{2}}\left( \frac{\dot{N}}{N}+\frac{\dot{\gamma}}{\gamma }%
-3NH\right) \right] ,
\end{equation}%
and the field equations are%
\begin{gather}
3H^{2}f^{\prime }(Q)+\frac{1}{2}\left( f(Q)-Qf^{\prime }(Q)\right) -\frac{%
3\gamma \dot{Q}f^{\prime \prime }(Q)}{2a^{2}}+3k\left( \frac{f^{\prime }(Q)}{%
a^{2}}-\frac{\dot{Q}f^{\prime \prime }(Q)}{2\gamma N^{2}}\right)  =  \rho _{m0}a^{-3}, \\
-\frac{2}{N}\frac{d}{dt}\left( f^{\prime }(Q)H\right) -3H^{2}f^{\prime }(Q)-%
\frac{1}{2}\left( f(Q)-Qf^{\prime }(Q)\right) +\frac{\gamma \dot{Q}f^{\prime
\prime }(Q)}{2a^{2}}-k\left( \frac{f^{\prime }(Q)}{a^{2}}+\frac{3\dot{Q}%
f^{\prime \prime }(Q)}{2\gamma N^{2}}\right) = 0, \\
\dot{Q}f^{\prime \prime \prime }\left( Q\right) \left( 1+\frac{ka^{2}}{%
N^{2}\gamma ^{2}}\right) +f^{\prime \prime }\left( Q\right) \left( \ddot{Q}%
\left( 1+\frac{ka^{2}}{N^{2}\gamma ^{2}}\right) +\dot{Q}\left( \left( 1+%
\frac{ka^{2}}{N^{2}\gamma ^{2}}\right) NH+\left( 1-\frac{ka^{2}}{N^{2}\gamma
^{2}}\right) \frac{\dot{N}}{N}+\frac{2\dot{\gamma}}{\gamma }\right) \right) = 0.
\end{gather}

In order to understand the dynamical properties and the degrees of freedom
for each set of the field equations, in the following we present the
minisuperspace description for the above cosmological models.

\subsection{Minisuperspace description}

Minisuperspace description is important in gravitational physics, because it
allows us to understand the dynamics and give physical description on the
geometrodynamical terms. In the minisuperspace description the degrees of
freedom provided by the geometrodynamical components can be attributed to
scalar fields. The latter can be used for the construction of a point-like
Lagrangian. The existence of a point-like Lagrangian, which means that the
field equations follows from a variation principle, is essential in order to apply Noether's theorems for the derivation of the variational symmetries
and of the corresponding conservation laws.

In the following lines we present four point-like Lagrangian functions which
describe the gravitational field equations for the $f\left( Q\right) $%
-cosmology for the four families of connections. In what it follows we
consider the scalar field $\phi $ defined as \cite{mini} 
\begin{eqnarray}
\phi &=&f^{\prime }\left( Q\right) ,~  \label{cl.1} \\
V\left( \phi \right) &=&\left( f(Q)-Qf^{\prime }(Q)\right) .  \label{cl.2}
\end{eqnarray}

For the connection $\Gamma ^{A},$ the point-like Lagrangian is 
\begin{equation}
L\left( \Gamma ^{A}\right) =-\frac{3}{N}\phi a\dot{a}^{2}+\frac{1}{2}%
Na^{3}V\left( \phi \right) -N\rho _{m0}\text{.}  \label{lg1}
\end{equation}%
Notice that there is no derivative of the scalar field $\phi$ in Lagrangian (\ref{lg1}). 

The field equations for the connections $\Gamma ^{B}$ are described by the
point-like Lagrangian 
\begin{equation}
L\left( \Gamma ^{B}\right) =-\frac{3}{N}\phi a\dot{a}^{2}-\frac{3}{2N}a^{3}%
\dot{\phi}\dot{\psi}+\frac{N}{2}a^{3}V\left( \phi \right) -N\rho _{m0}~\text{%
where}~\dot{\psi}=\gamma \text{.}  \label{lg2}
\end{equation}%
The scalar $\gamma $ is a scalar field with kinetic term in the Lagrangian
function, and the equation of motion for the scalar field $\psi $, is the
constraint equation for the connection.

In addition, for the connections $\Gamma ^{C}$ and $\Gamma ^{D}$ the
corresponding point-like Lagrangians are 
\begin{equation}
L\left( \Gamma ^{C}\right) =-\frac{3}{N}\phi a\dot{a}^{2}-\frac{3}{2}aN\frac{%
\dot{\phi}}{\dot{\Psi}}+\frac{N}{2}a^{3}V\left( \phi \right) -N\rho _{m0}~%
\text{where}~\dot{\Psi}=\frac{1}{\gamma },  \label{lg3}
\end{equation}%
and 
\begin{equation}
L\left( \Gamma ^{D}\right) =-\frac{3}{N}\phi a\dot{a}^{2}+\frac{3}{2N}ka^{3}%
\dot{\phi}\dot{\Psi}-\frac{3}{2}aN\frac{\dot{\phi}}{\dot{\Psi}}+3k\phi aN+%
\frac{N}{2}a^{3}V\left( \phi \right) -N\rho _{m0}~\text{where}~\dot{\Psi}=%
\frac{1}{\gamma }.  \label{lg4}
\end{equation}%
We remark that the new scalar field $\Psi $ which attributes the dynamical
degrees of freedom introduced by the connection does not have a canonical
kinetic term. Kinetic terms of this form are well known in fluid dynamics in
dynamical systems which describe the shallow-water equations in the Lagrange
variables.

In the following we shall consider the lapse function to be a constant, i.e. 
$N=1$. In this case, the constraint equations can be seen, as the
conservation law of the Hamiltonian for the gravitational field equations.
Moreover, by definition, $V\left( \phi \right) $ should not be a constant,
or a linear function, otherwise $f\left( Q\right) $ will be linear, and we want to focus on deviations from GR.

\section{Variational Symmetries and Conservation laws}

\label{sec3}

Symmetry analysis is a powerful approach for the analysis and the study of
nonlinear partial differential equations. It was
established by S. Lie \cite{noe1} and it is based on the existence of
transformations which keep the given dynamical system invariant. The
existence of a sufficient number of suitable symmetries enables the equation
to be solved through successive reductions of order, a series of
quadratures, or the determination of first integrals.

Consider the dynamical system described by the following set of second-order
differential equations 
\begin{equation}
\mathbf{\ddot{x}}=\mathbf{\Omega }\left( t,\mathbf{x},\mathbf{\dot{x}}%
\right) .  \label{Lie.0}
\end{equation}%
Moreover, we assume the system follows from a variational principle with
Lagrangian function 
\begin{equation}
L=L\left( t,\mathbf{x},\mathbf{\dot{x}}\right) .  \label{Lie.1}
\end{equation}

In the augmented space of dependent and independent variables $\{t,\mathbf{x}%
\}$ we define the infinitesimal transformation 
\begin{eqnarray}
\bar{t} &=&t+\varepsilon \xi \left( t,\mathbf{x}\right) , \\
\mathbf{\bar{x}} &=&\mathbf{x}+\mathbf{\eta }\left( t,\mathbf{x}\right) ,
\end{eqnarray}%
with generator the vector field 
\begin{equation}
\mathbf{X}=\xi \left( t,\mathbf{x}\right) \partial _{t}+\mathbf{\eta }\left(
t,\mathbf{x}\right) \partial _{\mathbf{x}}.
\end{equation}
The dynamical system (\ref{Lie.0}) remain invariant under the action of the
latter infinitesimal transformation if and only if \cite{noe2} 
\begin{equation}
\mathbf{X}^{\left[ 2\right] }\left( \mathbf{\ddot{x}}-\mathbf{\Omega }\left(
t,\mathbf{x},\mathbf{\dot{x}}\right) \right) =0,
\end{equation}%
where $\mathbf{X}^{\left[ 2\right] }$ is the second extension of the vector $%
X$ in the jet buddle $\left\{ t,\mathbf{x},\mathbf{\dot{x}},\mathbf{\ddot{x}}%
\right\} $, defined as%
\begin{equation}
\mathbf{X}^{\left[ 2\right] }=\mathbf{X}+\mathbf{\eta }^{\left[ 1\right]
}\partial _{\dot{x}}+\mathbf{\eta }^{\left[ 2\right] }\partial _{\ddot{x}},
\end{equation}%
where
\begin{equation}
\mathbf{\eta }^{\left[ 1\right] } = \left( \mathbf{\dot{\eta}}-\mathbf{\dot{%
x}}\dot{\xi}\right) \quad , \quad \mathbf{\eta }^{\left[ 2\right] } = \left( \mathbf{\dot{\eta}}^{\left[ 1 \right] }-\mathbf{\ddot{x}}\dot{\xi}\right) \,.
\end{equation}

Lie symmetries can be applied in many ways for the analysis of the dynamical
system. However, the most important use of the Lie symmetries is that they
can be used to determine conservation laws. The pioneer work of E. Noether \cite{noe0} provides one of the most
systematic ways for the construction of conservation laws. It provides an one-to-one relation between the variational symmetries and the respective conservation laws. Noether's work is summarized in two theorems. The first
theorem gives a simple algebraic relation that can be used to calculate the one-parameter point transformations that leave the variation of the action integral invariant. On the other hand, the second theorem relates the variational symmetries to the admitted conservation laws. Because variational symmetries leave the dynamical system invariant, it means that
Noether symmetries are also Lie symmetries, but the inverse is not necessarily true. Only some of the Lie symmetries are Noether symmetries \cite{noe1}.

Noether's first theorem states, that if $X$ is a Lie symmetry for the
dynamical system (\ref{Lie.0}) then there exist a function $g$, such that
the following algebraic condition is true \cite{noe2}%
\begin{equation}
\mathbf{X}^{\left[ 1\right] }L+L\dot{\xi}=\dot{g}.  \label{ns.01}
\end{equation}%
When the latter is true, Noether's second theorem states that the function 
\begin{equation}
I\left( t,\mathbf{x,\dot{x}}\right) =\xi \mathcal{H}-\frac{\partial L}{%
\partial \mathbf{\dot{x}}}\mathbf{\dot{x}}+g,  \label{ns.02}
\end{equation}%
is a conservation law, where 
\begin{equation}
\mathcal{H}=\frac{\partial L}{\partial \mathbf{\dot{x}}}\mathbf{\dot{x}}-L,
\end{equation}%
and $\mathcal{H}$ is the Hamiltonian function.

\section{Classification of Noether Symmetries in $f\left( Q\right) -$%
cosmology}

\label{sec4}

The modern treatment of the classification problem was established by Ovsiannikov \cite{ovsi}.
Specifically, the solution of the symmetry classification problem provides
constraints for the unknown functions of the dynamical system, such that
symmetries to emerge. Within the gravitational theory and cosmology, the variational symmetries
have been used to classify the unknown functions and parameters for the
proposed gravitational models. This approach is two-fold. It allows us to
determine the free parameters and functions of the model where conservation
laws exist such that exact and analytic solutions to exist. Moreover, the
application of the Noether symmetry condition is a geometric selection rule
in agreement with the geometric characteristics of gravity \cite{anb2}.

In particular, the Noether symmetries are generated by the geometric
characteristics of the minisuperspace, thus, when we impose the existence of
Noether symmetries, the given gravitational model self-provide the
constraints for the free functions and parameters. For more details we refer
in the discussion in \cite{anb2}.

For the four different Lagrangian functions of $f\left( Q\right) -$gravity,
we employ the Noether symmetry condition (\ref{ns.01}) to determine the
functional form of the potential $V\left( \phi \right) $, that is, of the $%
f\left( Q\right) $ function such that Noether symmetries and Noether conservation laws exist. The latter are used to derive exact and analytic
solutions for the field equations.

\subsection{Connection $\Gamma ^{A}$}

For the connection $\Gamma ^{A}$ and Lagrangian function (\ref{lg1}),
Noether symmetries appears only for the power-law potential 
\begin{equation}
V\left( \phi \right) =V_{0}\phi ^{n}.
\end{equation}
The admitted symmetry vectors are%
\begin{eqnarray*}
X_{1} &=&\partial _{t}. \\
X_{2} &=&a^{\frac{3}{2n}-\frac{1}{2}}\partial _{a}-\frac{3}{n}\phi a^{\frac{%
3(1-n)}{2n}}\partial _{\phi }, \\
X_{3} &=&\frac{\left( n-1\right) }{2}t\partial _{t}+\frac{n}{3}a\partial
_{a}-\phi \partial _{\phi },
\end{eqnarray*}%
and the corresponding conservation laws follow from Noether's second theorem (%
\ref{ns.02}), 
\begin{eqnarray}
I\left( X_{1}\right) &=&3\phi a\dot{a}^{2}+\frac{1}{2}a^{3}V\left( \phi
\right) -\rho _{m0}\equiv 0, \\
I\left( X_{2}\right) &=&a^{\frac{3}{2n}+\frac{1}{2}}\phi \dot{a}, \\
I\left( X_{3}\right) &=&\frac{\left( n-1\right) }{2}t\rho _{m0}+2n\phi a^{2}%
\dot{a}.
\end{eqnarray}
We remind that for arbitrary potential $V\left( \phi \right) $, only the
trivial symmetry $X_{1}$ occurs, with conservation law the constraint
equation.

\subsection{Connection $\Gamma ^{B}$}

The Lagrangian function (\ref{lg2}) for arbitrary potential $V\left( \phi
\right) $ admits as Noether symmetries the vector fields%
\begin{equation*}
Y_{1} = \partial _{t}\quad , \quad Y_{2} = \partial _{\psi }\,,
\end{equation*}%
with conservation laws%
\begin{eqnarray*}
I\left( Y_{1}\right) &=&3\phi a\dot{a}^{2}+\frac{3}{2}a^{3}\dot{\phi}\dot{%
\psi}+\frac{1}{2}a^{3}V\left( \phi \right) -\rho _{m0}\equiv 0, \\
I\left( Y_{2}\right) &=&a^{3}\dot{\phi}.
\end{eqnarray*}
Furthermore, for the power-law potential $V\left( \phi \right) =V_{0}\phi
^{n}$ the additional Noether symmetry occurs%
\begin{equation*}
Y_{3}=\left( n-1\right) t\partial _{t}+\frac{2}{3}\left( n+1\right) a+\phi
\partial _{\phi },
\end{equation*}
with conservation law%
\begin{equation*}
I\left( Y_{3}\right) =\left( n-1\right) t\rho _{m0}-4\left( n+1\right) \phi
a^{2}\dot{a}+3\phi a^{3}\dot{\psi}.
\end{equation*}%
\bigskip
As mentioned above, in the next section, we will utilize the above conservation laws in order to find cosmological solutions.

\subsection{Connection $\Gamma ^{C}$}

For the third connection of the spatially flat FLRW geometry, namely
connection $\Gamma ^{C}$, the point-like Lagrangian (\ref{lg3}) for
arbitrary potential $V\left( \phi \right) $ admits the Noether symmetries%
\begin{equation*}
    Z_{1} = \partial _{t}\quad , \quad Z_{2}  = \partial _{\Psi }\,,
\end{equation*}
with conservation laws%
\begin{eqnarray}
I\left( Z_{1}\right) &=&3\phi a\dot{a}^{2}-\frac{3}{2}a\frac{\dot{\phi}}{%
\dot{\Psi}}+\frac{1}{2}a^{3}V\left( \phi \right) -\rho _{m0}\equiv 0, \\
I\left( Z_{2}\right) &=&a\frac{\dot{\phi}}{\dot{\Psi}^{2}}.
\end{eqnarray}
However, for the power-law potential $V\left( \phi \right) =V_{0}\phi ^{n}$
there exists the additional symmetry vector%
\begin{equation}
Z_{3}=2(n-1)\partial _{t}+\left( n+1\right) a\partial _{a}-\phi \partial
_{\phi }+\frac{2}{3}\left( n-2\right) \psi \partial _{\psi }\,,
\end{equation}%
with conservation law%
\begin{equation}
I\left( Z_{3}\right) =2(n-1)t\rho _{m0}+\left( n+1\right) a^{2}\phi \dot{a}-%
\frac{3}{2}a\frac{\phi }{\dot{\Psi}}-\left( n-2\right) a\psi \frac{\dot{\phi}%
}{\dot{\Psi}^{2}}.
\end{equation}

\subsection{Connection $\Gamma ^{D}$}

Last but not least, the Lagrangian function (\ref{lg3}) for arbitrary potential function $V\left(
\phi \right) $, admits the Noether symmetries 
\begin{equation*}
W_{1} =\partial _{t}\quad , \quad W_{2} = \partial _{\Psi }\,,    
\end{equation*}
with conservation laws%
\begin{eqnarray}
I\left( W_{1}\right) &=&3\phi a\dot{a}^{2}+\frac{a^{3}}{2}V\left( \phi
\right) -\frac{3\dot{\phi}}{2a^{2}\dot{\Psi}}+3ka^{3}\left( \frac{\phi }{%
a^{2}}-\frac{\dot{\phi}\dot{\Psi}}{2}\right) -\rho _{m0}\equiv 0, \\
I\left( W_{2}\right) &=&k\dot{\phi}+a\frac{\dot{\phi}}{\dot{\Psi}^{2}}\,,
\end{eqnarray}
while for the quadratic potential $V\left( \phi \right) =V_{0}\phi ^{2}$, the dynamical system admits the additional Noether symmetry%
\begin{equation*}
W_{3}=t\partial _{t}-a\partial _{a}+2\phi \partial _{\phi },
\end{equation*}%
with conservation law%
\begin{equation}
I\left( W_{3}\right) =t\rho _{m0}-6\phi a^{2}\dot{a}+\frac{3a\phi }{\dot{\Psi%
}}-3ka^{3}\phi \dot{\Psi}.
\end{equation}

Before we continue with the application of the above conservation laws, we
should derive the corresponding $f\left( Q\right) $ function which
corresponds to the power-law potential $V\left( \phi \right) =V_{0}\phi ^{n}$%
. Thus, from (\ref{cl.1}) and (\ref{cl.2}) we define the Clairaut equation 
\begin{equation}
\left( f(Q)-Qf^{\prime }(Q)\right) =\left( V_{0}f^{\prime }\left( Q\right)
\right) ^{n},
\end{equation}%
with nonlinear solution the power-law function 
\begin{equation}
f\left( Q\right) \simeq Q^{\frac{n}{n-1}}\,.
\end{equation}
This verifies that no other functional form of $f(Q)$, beyond the power-law one is invariant under point transformations and thus admits Noether symmetries.

\section{Cosmological solutions}

\label{sec5}

In this Section we utilize the Noether symmetries to determine exact
solutions for the field equations. By definition a $2d-$dimensional
Hamiltonian system will be characterized as Liouville integrable, if and
only if, there exist $d$ conservation laws which are independent and in
involution. From the results of the previous section we conclude the field
equations for connection $\Gamma ^{A}$ are Liouville integrable, while the
field equations for the connection $\Gamma ^{B}$ are integrable only when $%
\rho _{m0}=0$. The other two cosmological models derived by the connections $%
\Gamma ^{C}$ and $\Gamma ^{D}$ are not integrable by the variational point
symmetries. Nevertheless, the symmetry vectors are used to determine
invariant functions and to calculate exact solutions, known as similarity
solutions.

\subsection{Connection $\Gamma ^{A}$}

For the first connection and for the power-law potential we calculate the
analytic solution 
\begin{eqnarray}
a\left( t\right) &=&a_{0}\left( t-t_{0}\right) ^{\frac{2n}{3\left(
n-1\right) }},~ \\
\phi \left( t\right) &=&\left( \frac{\left( n-1\right) V_{0}}{4}t^{2}\right)
^{\frac{1}{1-n}}
\end{eqnarray}%
with constraint 
\begin{equation}
\rho _{m0}=2^{2+\frac{3}{n-1}}a_{0}^{3}\left( \left( 3-\frac{3}{n}\right)
\left( n-1\right) \right) ^{\frac{n}{1-n}}V_{0}^{\frac{1}{1-n}}\left(
1+n\right) .
\end{equation}%
This is the general solution for the cosmological model.

\subsection{Connection $\Gamma ^{B}$}

For the field equations for the connection $\Gamma ^{B}$ we define the
momentum as%
\begin{equation}
\dot{a} =-\frac{p_{a}}{6a\phi }\,, \quad
\dot{\phi} = -\frac{2p_{\psi }}{3a^{3}}\,, \quad
\dot{\psi} = -\frac{2p_{\phi }}{3a^{3}}\,,    
\end{equation}
such that the Hamiltonian function and the conservation laws read%
\begin{eqnarray}
\mathcal{H} &=&\frac{p_{a}^{2}}{12a\phi }+\frac{2}{3a^{3}}p_{\phi }p_{\psi }+%
\frac{V_{0}}{2}a^{3}\phi ^{n}\equiv 0, \\
I\left( Y_{2}\right) &=&-\frac{2}{3}p_{\psi }\equiv I_{1}, \\
I\left( Y_{3}\right) &=&\frac{\left( n+1\right) }{3}ap_{a}-2\phi p_{\phi
}\equiv I_{2}.
\end{eqnarray}
Consequently, it follows that the generalized the generalized momenta are
\begin{eqnarray}
p_{a} &=&\frac{\left( n+1\right) I_{1}-\sqrt{I_{1}^{2}\left( n+1\right)
^{2}-6\left( I_{1}I_{2}+V_{0}a^{6}\phi ^{n+1}\right) }}{a}, \\
p_{\phi } &=&\frac{I_{1}\left( n+1\right) ^{2}-3I_{2}-\sqrt{\left(
n+1\right) ^{2}I_{1}^{2}\left( n+1\right) ^{2}-6\left(
I_{1}I_{2}+V_{0}a^{6}\phi ^{n+1}\right) }}{\phi }, \\
p_{\psi } &=&-\frac{3}{2}I_{1}.
\end{eqnarray}%
and the field equations reduce to a set of first order differential equations that read
\begin{eqnarray}
\dot{a} &=&-\frac{\left( n+1\right) I_{1}-\sqrt{I_{1}^{2}\left( n+1\right)
^{2}-6\left( I_{1}I_{2}+V_{0}a^{6}\phi ^{n+1}\right) }}{6a^{2}\phi },
\label{ll01} \\
\dot{\phi} &=&\frac{I_{1}}{a^{3}},  \label{ll02} \\
\dot{\psi} &=&-\frac{2}{3}\frac{I_{1}\left( n+1\right) ^{2}-3I_{2}-\sqrt{%
\left( n+1\right) ^{2}\left( I_{1}^{2}\left( n+1\right) ^{2}-6\left(
I_{1}I_{2}+V_{0}a^{6}\phi ^{n+1}\right) \right) }}{a^{3}\phi }.
\end{eqnarray}

In the special limit where $I_{2}=\frac{1}{6}I_{1}\left( 1+n\right) ^{2}$
the evolution of the scale factor and the scalar field read%
\begin{eqnarray}
\dot{a} &=&-\frac{\left( n+1\right) I_{1}-\sqrt{-6V_{0}a^{6}\phi ^{n+1}}}{%
6a^{2}\phi },  \label{ll.03} \\
\dot{\phi} &=&\frac{I_{1}}{a^{3}},  \label{ll.04}
\end{eqnarray}%
From the above, it follows that
\begin{equation}
\frac{d\phi }{da}=-\frac{I_{1}\phi }{a\left( I_{1}\left( n+1\right) +\sqrt{%
-6V_{0}}a^{3}\phi ^{\frac{1+n}{2}}\right) },
\end{equation}%
which admits the solution%
\begin{equation}
\ln \phi +\frac{2}{5\left( n+1\right) }\ln \left( \frac{I_{1}\left(
n+1\right) \sqrt{-V_{0}}a^{3}\phi ^{\frac{1+n}{2}}}{6\sqrt{-6V_{0}}a^{3}\phi
^{\frac{1+n}{2}}-5I_{1}\left( n+1\right) }\right) =\ln \phi _{0}.
\end{equation}%
Thus, for large values of $a^{3}\phi ^{\frac{1+n}{2}}$, we have
\begin{equation}
\ln \phi \simeq const.
\end{equation}%
which means that the asymptotic solution is that of the de Sitter universe.
This is in agreement with the phase-space analysis performed in \cite%
{andynfq}. In particular, in \cite{andynfq} it was found that there exist a set of
initial conditions where the unique attractor for the model is the de Sitter
universe. In the appearance of a matter source the same result is also true 
\cite{andynfq2}. Consequently, this power-law theory can explain the recent
cosmic acceleration of the universe.

\begin{figure}[tbph]
\centering\includegraphics[width=1\textwidth]{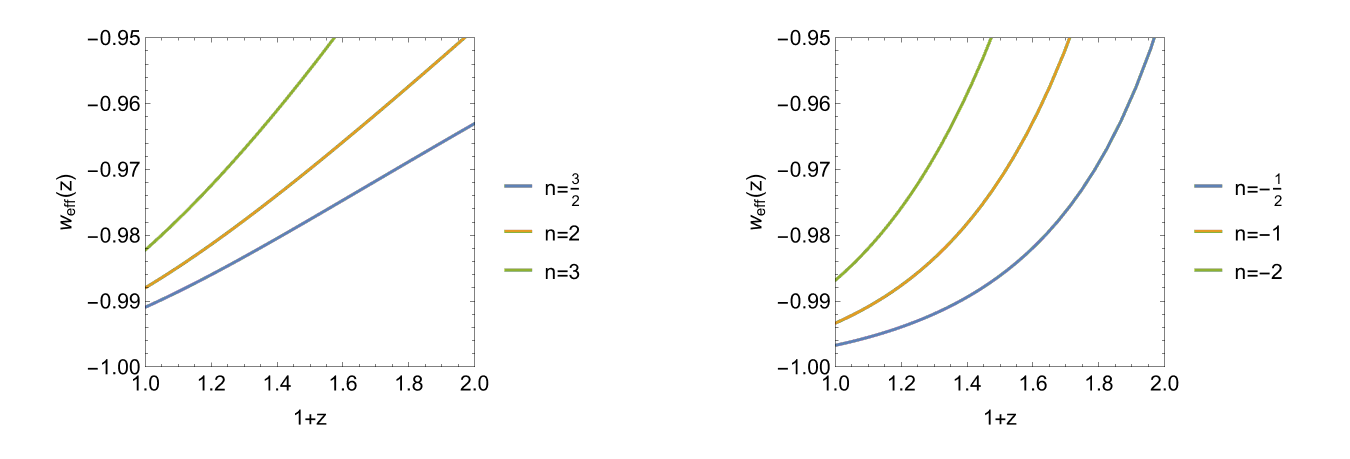}
\caption{Qualitative evolution of the effective equation of state parameter $%
w_{eff}\left( z\right) $ as it is given by the numerical simulations of the
dynamical system (\protect\ref{ll.03}), (\protect\ref{ll.04}) where $z$ is
the redshift, $1+z=a^{-1}$. Plots are for initial conditions We have assumed
the initial conditions $t_{0}$ is the present time, $a\left( t_{0}\right) =1$%
,~$\protect\phi \left( t_{0}\right) =1\,$. Left Fig. is for $I_{1}=0.01$,~$%
\protect\sqrt{-6V_{0}}=0.2$, while Right Fig. is for $I_{1}=-0.01$ and $%
\protect\sqrt{-6V_{0}}=0.2$. The lines are for different values of parameter 
$n$ as they are given in the legends. }
\label{fig1}
\end{figure}

In Fig. \ref{fig1}, we present the qualitative evolution of the effective
equation of state parameter $w_{eff}=-1-\frac{2}{3}\frac{\dot{H}}{H^{2}},\,\ 
$as it is given from numerical simulations for the dynamical system (\ref%
{ll.03}) and (\ref{ll.04}). 

\subsection{Connection $\Gamma ^{C}$}

For the field equations of the third connection $\Gamma ^{C}$, and from the
symmetry vector $Z_{3}$ we determine the similarity solution%
\begin{eqnarray}
a\left( t\right) &=&a_{0}\left( t-t_{0}\right) ^{p_{1}}, \\
\phi \left( t\right) &=&\phi _{0}\left( t-t_{0}\right) ^{p_{2}}, \\
\Psi \left( t\right) &=&\Psi _{0}\left( t-t_{0}\right) ^{p_{3}},
\end{eqnarray}%
where%
\begin{eqnarray}
p_{1} &=&\frac{3n-1}{5\left( n-1\right) },~ \\
p_{2} &=&-\frac{2}{n-1},~ \\
p_{3} &=&\frac{4\left( n-2\right) }{5\left( n-1\right) },
\end{eqnarray}%
with constraints $\rho _{m0}=0$ and 
\begin{eqnarray}
V_{0} &=&\frac{12}{25}\phi _{0}^{1-n}\left( 3n^{2}+8n-3\right) , \\
\Psi _{0} &=&-\frac{25}{4a_{0}^{2}}\frac{\left( n-1\right) ^{2}}{\left(
3n^{3}-4n^{2}-5n+2\right) }.
\end{eqnarray}
This is a new exact cosmological solution that has not been reported so far. 

\subsection{Connection $\Gamma ^{D}$}

Finally, for the case with nonzero spatially curvature and connection~$%
\Gamma ^{D}$, we derive the similarity solution%
\begin{eqnarray}
a\left( t\right) &=&a_{0}\left( t-t_{0}\right) ~, \\
\phi \left( t\right) &=&\phi _{0}\left( t-t_{0}\right) ^{-1}, \\
\Psi \left( t\right) &=&\Psi _{0}\ln \left( t-t_{0}\right) ,
\end{eqnarray}%
in which $\rho _{m0}=0$ and%
\begin{eqnarray}
\Psi _{0} &=&-\frac{1}{3a_{0}^{2}}, \\
V_{0} &=&\frac{4}{\phi _{0}}\left( 3-\frac{k}{a_{0}^{2}}\right) .
\end{eqnarray}%
These results are in agreement with the scaling solutions determined in \cite%
{ndim1}. It is interesting to mention that the existence of the scaling
solutions is related with the appearance of the Noether symmetry vectors.

\section{Conclusion}

\label{sec6}
The Noether symmetry analysis was applied to the field equations of symmetric teleparallel $f(Q)-$cosmology within an FLRW geometry. These field equations exhibit the property of admitting a minisuperspace description. In fact, there are four distinct families of connections for the FLRW geometry within symmetric teleparallel theory, each of which leads to a different minisuperspace Lagrangian.

By applying Noether's theorem, we constrained the free functions and parameters of the four gravitational Lagrangian functions. The Noether symmetry condition led to the conclusion that the only function $f(Q)$ for which the field equations possess point symmetries is the power-law model. This result implies that, for any arbitrary $f(Q)-$theory, when the power-law components dominate, the gravitational field equations admit Noetherian conservation laws. These conservation laws can then be used to determine exact and analytic solutions.

This analysis opens new directions for exploring the integrability properties within the symmetric teleparallel theory of gravity, offering a promising framework for further investigation into conserved quantities and the exact solutions of $f(Q)$-gravity.

\begin{acknowledgments}
This paper is based upon work from COST Action CA21136 \textit{Addressing
observational tensions in cosmology with systematics and fundamental physics}
(CosmoVerse) supported by COST (European Cooperation in Science and
Technology). The work of KD supported by the PNRR-III-C9-2022 call, with
project number 760016/27.01.2023. GL\ \& AP thanks the support of VRIDT
through Resoluci\'{o}n VRIDT No. 096/2022 and Resoluci\'{o}n VRIDT No.
098/2022. This study was supported by FONDECYT 1240514, Etapa 2024. AP
thanks the Woxsen University for the hospitality provided while part of this
work was carried out. K.F.D. was supported by the PNRR-III-C9-2022–I9 call, with project number 760016/27.01.2023.
\end{acknowledgments}


\end{document}